\documentclass[twocolumn,english]{article}

\usepackage[affil-it]{authblk}
\usepackage{amsmath}
\usepackage{amssymb}
\usepackage[T1]{fontenc}
\usepackage[latin9]{inputenc}
\usepackage{refstyle}
\usepackage{color,hyperref}
    \catcode`\_=11\relax
    \newcommand\email[1]{\_email #1\q_nil}
    \def\_email#1@#2\q_nil{%
      \href{mailto:#1@#2}{{\emailfont #1\emailampersat #2}}
    }
    \newcommand\emailfont{\sffamily}
    \newcommand\emailampersat{{\color{red}\small@}}
    \catcode`\_=8\relax

\makeatletter

\RS@ifundefined{subref}
  {\def\RSsubtxt{section~}\newref{sub}{name = \RSsubtxt}}
  {}
\RS@ifundefined{thmref}
  {\def\RSthmtxt{theorem~}\newref{thm}{name = \RSthmtxt}}
  
  {}
\RS@ifundefined{lemref}
  {\def\RSlemtxt{lemma~}\newref{lem}{name = \RSlemtxt}}
  {}

\usepackage{amsmath}

\makeatother

\usepackage{babel}
\usepackage{mathrsfs}
\begin{document}
\title{Phase space ensembles for classical and quantum-classical systems. }
\author{A. D. Berm\'udez Manjarres}
\affil{\footnotesize Universidad Distrital Francisco Jos\'e de Caldas\\ Cra 7 No. 40B-53, Bogot\'a, Colombia\\ \email{ad.bermudez168@uniandes.edu.co}}
\twocolumn[
  \begin{@twocolumnfalse}

 \maketitle

\centering
    \begin{minipage}{.9\textwidth}
\begin{abstract}
We develop a so-called theory of ensembles in phase space and use
it to investigate the construction of a quantum-classical hybrid theory.
We use Galilei covariance and the Lie algebra of the Galilei group
as a guide to constructing the hybrid model presented here. In particular,
we chose the interaction term between the classical and the quantum
sector so that the equations are Galilei covariant. Our approach points
out a possible connection between two previously unrelated hybrid
systems.
\end{abstract}
\hspace{2cm}
    \end{minipage}
\end{@twocolumnfalse}
]
\section{Introduction}

There have been multiple attempts to create a consistent theory for
coupled classical and quantum systems (see \cite{barcelo} and references
therein). The reasons for this endeavor are multi-fold and varied.
They include a better description of mesoscopic systems \cite{barcelo},
the measurement problem of quantum mechanics \cite{sudarshan1}, and
the interaction of a quantum system with classical gravity \cite{gravity}.

Quantum-classical hybrid theories tend to follow the same pattern,
they first try to put classical and quantum mechanics on the same
mathematical framework and then make them interact somehow. There
are two approaches of relevance for this paper. The first one is based
on the Koopman von Neumann (KvN) operational version of classical
mechanics \cite{koopman,koopman2,koopman3}, and the idea behind it
is to put classical mechanics in the same mathematical language of
wave functions and Hilbert spaces as quantum mechanics. While these
kind of hybrids has been criticized \cite{terno1,terno2,salcedo1,salcedo2,galilean},
there is still promising research made on them \cite{tronci1,DGBT,tronci2,tronci3}.

The second hybrid model of our interest is based on ensembles on configuration
space developed by Hall and Reginatto \cite{reginato,reginatto2,reginatto3}.
The idea of this model is to rewrite the classical Hamilton-Jacobi
equation and the Schr\"{o}dinger equation using a functional approach.
Once they are in the same mathematical setting, they can be easily
combined into a quantum-classical theory.

In this paper, we will rewrite the so-called Koopman-van Hove (KvH)
version of classical mechanics \cite{tronci1,DGBT} using Hall and
Reginatto functional approach. This results in a theory of ensembles
in phase space. We then construct a quantum-classical hybrid theory
using the symmetry principle of Galilean covariance as our guide.
It is shown that this approach leads to a set of equations resembling
Bondar, Tronci, and Gay-Balmaz hybrid equations \cite{tronci1,DGBT}.
We then explore a possible projection from phase space to configuration
space to explore a possible connection between the KvH and the Hall
and Reginatto hybrids theories.

This work is organized as follows: in section 2 we review the formalism
of ensembles in configuration space. We show how to obtain the Hamilton-Jacobi
and the Schr\"{o}dinger equations from a functional approach. In section
2.2 we give the definition of an observable in this formalism, and
in section 2.2 we show the explicit form of the observables/generators
of the space-time transformation of the Galilei group.

In section 3 we construct an ensemble in phase space. We show how
to obtain the KvH in a Hamiltonian setting. In section 3.2 we show
how to go back from a phase space description to configuration space,
thus recovering Hall and Reginatto classical ensembles.

In section 4 we construct a hybrid model, where the classical particle
is described in phase space and the quantum one is described in configuration
space. Galilean covariance dictates the possible interaction between
the classical and the quantum sectors in a non-trivial way. The above
results in a model that resemble the hybrid equations of Ref \cite{tronci1,DGBT}. 

\section{Ensembles on Configuration Space}

We want to describe the dynamics of a particle moving in three-dimensional
Euclidean space. Regardless of whether the particle is classical or
quantum, let us associate to it a density $\rho(\mathbf{x})$ that
tell us the probability of finding the particle around $\mathbf{x}$.
At all times, we demand that $\rho$ is positive and normalized
\begin{equation}
\int_{\mathbb{R}^{3}}d\mathbf{x}\,\rho=1.
\end{equation}
From now on, we will omit to specify the region of integration, and
we will understand that the integral is over the entirety of $\mathbb{R}^{3},$
or, eventually, some other appropriate space.

We postulate the existence of a function $S(\mathbf{x})$ that is
canonically conjugate to $\rho$. We demand that the dynamics of $\rho$
and $S$ is Hamiltonian
\begin{align}
\frac{\partial\rho}{\partial t} & =\frac{\delta\mathcal{H}}{\delta S},\\
\frac{\partial S}{\partial t} & =-\frac{\delta\mathcal{H}}{\delta\rho},
\end{align}
where the Form of the Hamiltonian functional $\mathcal{H}$ depends
on whether we are dealing with a classical or a quantum particle.
For a classical particle, the Hamiltonian functional

\begin{equation}
\mathcal{H}_{C}=\int d\mathbf{x}\,\rho\left(\frac{\left|\nabla S\right|^{2}}{2m}+V\right),\label{HC}
\end{equation}
leads to the Hamilton-Jacobi equation

\begin{equation}
\frac{\partial S}{\partial t}+\frac{\left|\nabla S\right|^{2}}{2m}+V=0,
\end{equation}
and the conservation of probability in configuration space
\begin{equation}
\frac{\partial\rho}{\partial t}+\nabla\cdot\left(\rho\frac{\nabla S}{m}\right)=0.\label{conservation}
\end{equation}
On the other hand, quantum dynamics is given by 
\begin{equation}
\mathcal{H}_{Q}=\mathcal{H}_{C}+\frac{\hbar^{2}}{2m}\int d\mathbf{x}\,\frac{\left|\nabla\rho\right|^{2}}{\rho}.\label{HQ}
\end{equation}
The Hamiltonian (\ref{HQ}) leads to the same conservation equation
(\ref{conservation}), but the equation for $S$ now reads

\begin{equation}
\frac{\partial S}{\partial t}+\frac{\left|\nabla S\right|^{2}}{2m}+V-\frac{\hbar^{2}}{2m}\frac{\nabla^{2}\sqrt{\rho}}{\sqrt{\rho}}=0.\label{SQ}
\end{equation}
The Schr\"{o}dinger equation is recovered from (\ref{SQ}), of course,
by the standard Madelung transformation $\psi=\sqrt{\rho}e^{iS/\hbar}$,
\begin{equation}
\frac{\partial\psi}{\partial t}=-\frac{\hbar^{2}}{2m}\nabla^{2}\psi+V\psi.
\end{equation}

Since $S$ is related to the phase of the Schr\"{o}dinger equation, from
now own we will call it the phase variable.

By working with a Hamiltonian functional, both classical and quantum
mechanics are written in the same mathematical framework. This allowed
Hall and Reginatto construct a quantum-classical hybrid theory
\cite{reginato,reginatto2,reginatto3}, we will show the equations
of that model in a later section.

Let us finish this section by pointing out that the equations of motion
for $\rho$ and $S$ can be written more compactly as 
\begin{align}
\frac{\partial\rho}{\partial t} & =\left\{ \rho,\mathcal{H}\right\} _{f},\\
\frac{\partial S}{\partial t} & =\left\{ S,\mathcal{H}\right\} _{f},
\end{align}
where the functional Poisson brackets is given by
\begin{equation}
\left\{ A,B\right\} _{f}=\int d\mathbf{x}\,\left(\frac{\delta A}{\delta\rho}\frac{\delta B}{\delta S}-\frac{\delta A}{\delta S}\frac{\delta B}{\delta\rho}\right).\label{functionalPB}
\end{equation}
Later we will deal with functional defined over functions over different
spaces. We will keep using the same symbol $\left\{ ,\right\} _{f}$
to denote Poisson brackets over configuration space, phase space,
or any other space and the context should indicate the domain of
integration.

\subsection{Observables}

By definition, in the ensemble formalism, an observable $A$ is a
functional of $\rho$ and $S$ that respects the conservation of the
probability and the positivity of $\rho$. These conditions are written
as 

\begin{align}
A[\rho,S+c] & =A[\rho,S]\;(\mathrm{c\,is\,a\,constant}),\nonumber \\
\frac{\delta A}{\delta S} & =0\;\mathrm{if}\,\rho(\mathbf{x})=0.
\end{align}
It follows in particular that only relative values and derivatives
of $S$ has an impact on the dynamics. A further requirement of homogeneity
is given by the requirement 
\begin{equation}
A[\lambda\rho,S]=\lambda A[\rho,S]
\end{equation}
where $\lambda$ is an arbitrary positive constant. 

See (\cite{reginato}) for the relationship between the observables
so defined and measured quantities in the laboratory.

For a classical ensemble, we can associate an observable to any function
of phase space $\alpha(\mathbf{x},\mathbf{p})$ by first recalling
that in the Hamilton-Jacobi theory, the momentum is given by the relation
$\mathbf{p}=\nabla S$. Then, we associate an observable related to
$f$ by

\begin{equation}
C_{\alpha}=\int d\mathbf{x}\,\rho\alpha(\mathbf{x},\nabla S).\label{classicalO}
\end{equation}
It can be shown that the Poisson bracket for classical observables
is isomorphic to the phase space Poisson bracket \cite{reginato}

\begin{equation}
\left\{ C_{\alpha},C_{\beta}\right\} _{f}=C_{\left\{ \alpha,\beta\right\} }.\label{iso1}
\end{equation}

Quantum observables are defined by the expectation value of the corresponding
operator

\begin{equation}
\mathcal{Q}_{\hat{M}}=\int d\mathbf{x}d\mathbf{x}'\,\sqrt{\rho\rho'}e^{i(S-S')/\hbar}\left\langle d\mathbf{x}'\right|\hat{M}\left|d\mathbf{x}\right\rangle .\label{quantumO}
\end{equation}
It follows that

\begin{equation}
\left\{ \mathcal{Q}_{\hat{M}},\mathcal{Q}_{\hat{N}}\right\} _{f}=\mathcal{Q}_{[\hat{M},\hat{N}]/i\hbar}.\label{iso2}
\end{equation}

With classical and quantum observables defined as above ``\emph{one
arrives at the non-trivial result that the algebras of each of these
two classes of observables are isomorphic to the algebras that arise
naturally in the phase space and Hilbert space representations of
classical and quantum mechanics}'' \cite{reginatto3}.

\subsection{Galilean symmetries and restrictions on composite systems}

For the position and momentum observables, both formulas (\ref{classicalO})
and (\ref{quantumO}) leads to the same functionals

\begin{align}
X_{i} & =\int d\mathbf{x}\,\rho x_{i},\\
\Pi_{j} & =\int d\mathbf{x}\,\rho\frac{\partial S}{\partial x_{j}}.
\end{align}
These two observables obey the canonical Poisson bracket relationship

\begin{equation}
\left\{ X_{i},\Pi_{j}\right\} _{f}=-\int d\mathbf{x}\,x_{i}\frac{\partial\rho}{\partial x_{j}}=\int d\mathbf{x}\,\rho\frac{\partial x_{i}}{\partial x_{j}}=\delta_{ij},\label{bracket}
\end{equation}
where we have used integration by parts to arrive at this result.
We stress out that the derivation of (\ref{bracket}) depends crucially
on $\rho$ going to zero fast enough at infinity, so there is no boundary
term when performing the integration by parts. Several of the Poisson
bracket results we will show later are derived in this way.

Now, notice that $\Pi$ has the dual role of being the momentum observable
and the generator of translations. Indeed, we can check the following
equation

\[
\rho(\mathbf{x}-\delta\mathbf{x})=\rho(\mathbf{x})-\delta\mathbf{x}\cdot\nabla\rho=\rho(\mathbf{x})+\delta\mathbf{x}\cdot\left\{ \rho,\Pi\right\} _{f}.
\]

Analogously, whether we use the classical or the quantum formula,
the angular momentum and the dynamic mass moment are given by the
functionals 

\begin{align}
L_{i} & =\int d\mathbf{x}\,\rho\left(\varepsilon_{ijk}x_{j}\frac{\partial S}{\partial x_{k}}\right),\\
G_{i} & =\int d\mathbf{x}\,\rho\left(mx_{i}-t\frac{\partial S}{\partial x_{i}}\right).
\end{align}
Just as for $\Pi$, the observables $L$ and $G$ also have a dual
role in this formalism. They are the generators of rotations and Galilean
boosts, respectively.

The functionals $\Pi,\:L$, and $G$ can be used to give a realization
of the Galilei algebra. As the generator of time translation, we can
use (\ref{HC}) or (\ref{HQ}), as long as we take $V=0$. These generators
obey the Lie algebra

\begin{align}
\left\{ \mathcal{H},\Pi_{i}\right\} _{f} & =0,\quad\left\{ \mathcal{H},L_{i}\right\} _{f}=0,\nonumber \\
\left\{ G_{i},\mathcal{H}\right\} _{f} & =\Pi_{i},\quad\left\{ L_{i},\Pi_{j}\right\} _{f}=\varepsilon_{ijk}\Pi_{k},\nonumber \\
\left\{ L_{i},L_{j}\right\}  & =\varepsilon_{ijk}L_{k},\quad\left\{ L_{i},G_{j}\right\} _{f}=\varepsilon_{ijk}G_{k},\nonumber \\
\left\{ \Pi_{i},\Pi_{j}\right\}  & =0,\quad\left\{ G_{i},\Pi_{j}\right\} _{f}=-m\delta_{ij},\nonumber \\
\left\{ G_{i},G_{j}\right\} _{f} & =0.\label{Galgebra}
\end{align}

Consider now two interacting particles (both classical, both quantum or one classical and the other quantum). The configuration space of
this system is $\mathbb{R}^{3}\times\mathbb{R}^{3}$. The probability
density and the action are now functions of the joint space, i.e., $\rho(\mathbf{x}_{1},\mathbf{x}_{2})$
and $S(\mathbf{x}_{1},\mathbf{x}_{2})$. A realization of the Galilei
algebra for the composite system is given by the functionals

\begin{align*}
\Pi_{i} & =\int d\mathbf{x}_{1}d\mathbf{x}_{2}\,\rho\left(\frac{\partial S}{\partial x_{1i}}+\frac{\partial S}{\partial x_{2i}}\right),\\
L_{i} & =\int d\mathbf{x}_{1}d\mathbf{x}_{2}\,\rho\left(\varepsilon_{ijk}x_{1j}\frac{\partial S}{\partial x_{1k}}+\varepsilon_{ijk}x_{2j}\frac{\partial S}{\partial x_{2k}}\right),\\
G_{i} & =\int d\mathbf{x}_{1}d\mathbf{x}_{2}\,\rho\left(mx_{1i}+mx_{2i}-t\frac{\partial S}{\partial x_{1i}}-t\frac{\partial S}{\partial x_{1i}}\right).
\end{align*}

The generator of time evolution is also the sum of the generator of
the subsystems, but here we allow an extra term that accounts for
the interaction.

\[
\mathcal{H}=\mathcal{H}_{1}+\mathcal{H}_{2}+\mathcal{H}_{int}.
\]
The question is, what properties should have $\mathcal{H}_{int}$
in order that $\mathcal{H}$ obeys the Galilei algebra? Clearly, $\mathcal{H}_{int}$
has to be invariant under translations, rotations, and Galilean boosts. 

\[
\left\{ \mathcal{H}_{int},\Pi_{i}\right\} =\left\{ \mathcal{H}_{int},L_{i}\right\} =\left\{ \mathcal{H}_{int},G_{i}\right\} =0.
\]
It can be checked that, for momentum-independent forces, $\mathcal{H}_{int}$
is restricted to functionals of the form
\begin{equation}
\mathcal{H}_{int}=\int d\mathbf{x}_{1}d\mathbf{x}_{2}\:\rho V(\left|\mathbf{x}_{1}-\mathbf{x}_{2}\right|).\label{Hint}
\end{equation}
This restriction imposed on $\mathcal{H}_{int}$ is independent on
whether $\mathcal{H}_{1}$ and $\mathcal{H}_{2}$ represent classical
or quantum particles.

\section{Ensembles on Phase Space}

In this section, we are going to take a functional ensemble approach to
classical mechanics in phase space. From now on, we will slightly
change our notation to indicate that the probability density and the
phase variable are now functions of phase-space coordinates. We designate
them by $\varrho(\mathbf{q},\mathbf{p})$ and $\mathcal{S}(\mathbf{q},\mathbf{p})$,
respectively.

As $\varrho$ is a probability density, it must be positive at all
times and normalized 

\begin{equation}
\int d\omega\,\varrho=1,
\end{equation}
where $d\omega=d\mathbf{q}d\mathbf{p}$ is the phase-space measure.

We must also have a phase-space conservation equation of the form
\begin{equation}
\frac{\partial\varrho}{\partial t}+\nabla_{\omega}\cdot\left(\varrho\dot{\omega}\right)=0,\label{ro1}
\end{equation}
were $\nabla_{\omega}$ is the phase-space gradient 

\[
\nabla_{\omega}=(\nabla_{q},\nabla_{p}).
\]
Using Hamilton equations, (\ref{ro1}) leads to Liouville equation 

\begin{equation}
\frac{\partial\mathcal{\varrho}}{\partial t}+\left\{ \varrho,H\right\} =0,\label{liouville}
\end{equation}
where the above bracket is the standard Poisson bracket of analytical
mechanics.

It seems there is some freedom in the choice in the equation of the
phase variable. A sensible choice is to identify $\mathcal{S}$ with
the action in phase space and demand that it obeys the equation

\begin{equation}
\frac{d\mathcal{S}}{dt}=\frac{\partial\mathcal{S}}{\partial t}+\left\{ \mathcal{S},H\right\} =\mathscr{L},\label{Sl}
\end{equation}
where $\mathscr{L}$ is the phase space Lagrangian of the system.
We make this choice because we later want to construct a quantum-classical
hybrid theory, hence our description of the classical system should
parallel quantum mechanics as closely as possible, and we know the relation
between phases and Lagrangian since the days of Dirac and Feynman. 

The equations (\ref{liouville}) and (\ref{Sl}) and the Madelung
transformation 
\begin{equation}
\psi(\mathbf{q},\mathbf{p})=\sqrt{\varrho}e^{i\mathcal{S}/\hbar},\label{phasemadelung}
\end{equation}
 lead to the recently named Koopman-van Hove equation \cite{DGBT,klein}

\[
\frac{\partial\psi}{\partial t}+\left\{ \mathcal{\psi},H\right\} =i\mathscr{L}.
\]

We restrict our attention to systems having a Hamiltonian and a Lagrangian
of the standard form

\begin{align}
H & =\frac{\mathbf{p}^{2}}{2m}+V(\mathbf{q}),\nonumber \\
\mathscr{L} & =\frac{\mathbf{p}^{2}}{2m}-V(\mathbf{q}).\label{Hamiltonian lagrangian}
\end{align}
Thus, we can write for the probability density and the phase variable
the following equations

\begin{align}
\frac{\partial\mathcal{\varrho}}{\partial t}+\nabla_{q}\varrho\cdot\frac{\mathbf{p}}{m}-\nabla_{p}\varrho\cdot\nabla_{q}V & =0,\label{ro2}\\
\frac{\partial\mathcal{S}}{\partial t}+\nabla_{q}\mathcal{S}\cdot\frac{\mathbf{p}}{m}-\nabla_{p}\mathcal{S}\cdot\nabla_{q}V & =\frac{\mathbf{p}^{2}}{2m}-V.\label{s2}
\end{align}

Equations (\ref{ro2}) and (\ref{s2}) are derivable from the Hamiltonian
functional

\begin{align}
\mathcal{H}_{C} & =\int d\omega\,\left[\mathcal{\varrho}(V-\frac{\mathbf{p}^{2}}{2m})+\frac{1}{2}\left(\mathcal{\varrho}\nabla_{q}\mathcal{S}-\mathcal{S}\nabla_{q}\mathcal{\varrho}\right)\cdot\frac{\mathbf{p}}{m}\right.\nonumber \\
 & \left.+\frac{1}{2}\left(\mathcal{S}\nabla_{p}\mathcal{\varrho}-\mathcal{\varrho}\nabla_{p}\mathcal{S}\right)\cdot\nabla_{q}V\right],\label{Hphasespace}
\end{align}

Since the position and the momentum work as coordinates in phase space,
we can proceed by analogy to section 1.2 and associate them the
following observables

\begin{align}
Q_{i} & =\int d\omega\,\mathcal{\varrho}q_{i},\label{position}\\
P_{j} & =\int d\omega\,\mathcal{\varrho}p_{j}.\label{momentum}
\end{align}
The generators of translation for the position and the momentum coordinates
can be respectively written, again by analogy, as

\begin{align}
A_{j} & =\int d\omega\,\mathcal{\varrho}\frac{\partial S}{\partial q_{j}},\label{translation}\\
\varLambda_{i} & =\int d\omega\,\mathcal{\varrho}\frac{\partial S}{\partial p_{i}}.
\end{align}
The following relations can be checked to be true

\[
\left\{ Q_{i},P_{j}\right\} _{f}=\left\{ Q_{i},\varLambda_{j}\right\} _{f}=\left\{ A_{j},P_{j}\right\} _{f}=\left\{ A_{j},\varLambda_{j}\right\} _{f}=0.
\]

\[
\left\{ Q_{i},A_{j}\right\} _{f}=\left\{ P_{i},\varLambda_{j}\right\} _{f}=\delta_{ij}.
\]

It is obvious from their definition, but it is worth remarking that
the momentum observable is different from the position translation
functional, $P_{j}\neq A_{j}$. This implies that the momentum will
not be part of the generators for the Galilei algebra. This is the
exact same situation that occur in the KvN \cite{me1,m2,GKVN1,GKVN2,GKVN3,GKVN4}
and the KvH \cite{m3} theories, where the self-adjoint operators
that generate the space-time transformations do not match the usual
physical quantities. 

Rotations and boosts are generated by 
\begin{align}
L_{i} & =\int d\omega\,\mathcal{\varrho}\varepsilon_{ijk}\left(q_{j}\frac{\partial S}{\partial q_{k}}+p_{j}\frac{\partial S}{\partial p_{k}}\right),\\
G_{i} & =\int d\omega\,\mathcal{\varrho}\left(mq_{i}-t\frac{\partial S}{\partial q_{i}}-m\frac{\partial S}{\partial p_{i}}\right).
\end{align}
The generators defined above have the desired effect on the position
and momentum observables

\begin{align}
\left\{ Q_{i},L_{j}\right\} _{f} & =\varepsilon_{ijk}Q_{k},\quad\left\{ P_{i},L_{j}\right\} _{f}=\varepsilon_{ijk}P_{k},\\
\left\{ Q_{i},G_{j}\right\} _{f} & =-t\delta_{ij},\quad\left\{ P_{i},G_{j}\right\} _{f}=-m\delta_{ij}.
\end{align}

These so defined $A,\,L,\,G$ and $\mathcal{H}_{C}$(with $V=0$)
obey the relations from Galilei algebra (\ref{Galgebra}).

For composite systems, we can proceed as in the previous section and
find that the allowed interaction term is of the form

\begin{equation}
\mathcal{H}_{int}=\int d\omega_{1}d\omega_{2}\:\rho V(\left|\mathbf{q}_{1}-\mathbf{q}_{2}\right|).\label{Hint2}
\end{equation}
Notice that the translation invariance of the interaction $(\left\{ A_{j},\mathcal{H}_{int}\right\} _{f}=0)$
is not the same as the condition of conservation of momentum. However,
it is true that (\ref{Hint2}) leads to a conserved momentum

\[
\left\{ P_{i},\mathcal{H}_{int}\right\} _{f}=0.
\]

Finally, notice that the (one particle) energy functional 
\begin{equation}
E=\int d\omega\,\mathcal{\varrho}\left(\frac{\mathbf{p}^{2}}{2m}+V(\mathbf{q})\right)\label{Energy functional}
\end{equation}
is completely different than the time evolution functional (\ref{Hphasespace}).

\subsection{Koopman origin of the observables in phase space}

Here we want to mention that in the KvN and the KvH operational formulations
of classical mechanics, the momentum and the position are commuting
self-adjoint operators, $[\hat{q}_{i},\hat{p}_{j}]=0,$ acting on
state vectors of the form 
\[
\left|\psi\right\rangle =\int\psi(\mathbf{q},\mathbf{p})\left|\mathbf{q},\mathbf{p}\right\rangle d\omega.
\]

On the other hand, the translation operator in the position coordinates
$\hat{\lambda}$ and the translation operator in the momentum coordinates
$\hat{\pi}$ are defined by the commutation relations

\[
[\hat{q}_{i},\hat{\lambda}_{j}]=[\hat{p}_{i},\hat{\pi}_{j}]=i\hbar\delta_{ij},
\]
and, when acting on wavefunctions, they have a derivative representation
of the form 

\begin{align*}
\hat{\lambda}_{j} & =-i\hbar\frac{\partial}{\partial q_{j}},\\
\hat{\pi}_{j} & =-i\hbar\frac{\partial}{\partial p_{i}}.
\end{align*}
The set $(\hat{q},\hat{p},\hat{\lambda},\hat{\pi})$ is irreducible
in the Hilbert space of the classical particle.

As the mathematical formalism behind quantum mechanics and the KvN
is the same (operators acting on a separable Hilbert space), we can
associate to any self-adjoint operator in the Koopman-von Neumann
theory an observable in our functional approach by a formula that
is analogous to (\ref{quantumO}), namely
\begin{equation}
\mathcal{Q}_{\hat{M}}=\int d\omega d\omega'\,\sqrt{\rho\rho'}e^{i(S-S')/\hbar}\left\langle \mathbf{q}',\mathbf{p}'\right|\hat{M}\left|\mathbf{q},\mathbf{p}\right\rangle .\label{quantumO-1}
\end{equation}
We must then have a phase space analogous to Eq.(\ref{iso2}). Hence,
the algebra of phase-space observables is isomorphic to the operator
algebra of the KvN theory.

Now, there are infinite ways to define a Hermitian operator in the
KvN theory from a function on phase space. We will mention three of
them, and we call them according to the convention given in \cite{m3}

\begin{align*}
f(\mathbf{q},\mathbf{p}) & \rightarrow f(\hat{\mathbf{q}},\hat{\mathbf{p}})\qquad\qquad\quad\quad(\mathrm{multiplication\:rule}),\\
f(\mathbf{q},\mathbf{p}) & \rightarrow\frac{\partial f}{\partial\mathbf{q}}\cdot\hat{\pi}+\frac{\partial f}{\partial\mathbf{p}}\cdot\hat{\lambda}\qquad\qquad\quad\quad(\mathrm{KvN\:rule}),\\
f(\mathbf{q},\mathbf{p}) & \rightarrow\frac{\partial f}{\partial\mathbf{q}}\cdot\hat{\pi}+\frac{\partial f}{\partial\mathbf{p}}\cdot\hat{\lambda}+f-\mathbf{p}\cdot\frac{\partial f}{\partial\mathbf{p}}\quad(\mathrm{KvH\:rule}).
\end{align*}

Notice that the operators formed by the multiplication rule form an Abelian subgroup of all possible operators in the KvN algebra. It
is impossible to describe the time evolution of the classical states
using only an Abelian algebra of operators \cite{bondar}. As an example
of functional that arise from operators defined by the multiplication
rule, we mention that the position (\ref{position}), the momentum
(\ref{momentum}), and the energy functionals (\ref{Energy functional})
are obtained by applying the formula (\ref{quantumO-1}) to the position
$\hat{q}_{i}$, the momentum $\hat{p}_{i}$, and the Hamiltonian operator
$\frac{\mathbf{\hat{p}}^{2}}{2m}+V(\hat{\mathbf{q}})$, respectively. 

The KvN rule is equivalent to the original prescription given by Koopman
and von-Neumann. The KvH rule is known as prequantization in the literature
of geometric quantization, and it has some advantages over the KvN
rule. For example, the KvH rule is one-to-one, while the KvN rule
is not injective. Additionally, in the KvH mechanics, the time evolution
is a unitary flow generated by the recently named covariant Liouvillian
operator
\begin{equation}
\mathcal{L}=\frac{\mathbf{\hat{p}}}{m}\cdot\hat{\pi}-\frac{\partial V}{\partial\mathbf{q}}\cdot\hat{\lambda}-\left(\frac{\mathbf{\hat{p}}^{2}}{2m}-V(\hat{\mathbf{q}})\right).\label{covariant liouvillian}
\end{equation}
This covariant Liouvillian is obtained by applying the KvH rule to
the Hamiltonian function (\ref{Hamiltonian lagrangian}). The Hamiltonian/time
evolution functional (\ref{Hphasespace}) is obtained from (\ref{covariant liouvillian})
using the formula (\ref{quantumO-1}).

Finally, it is worth remarking that the rules above all lead to operators
that are entirely different from each other, each with a different
spectrum and domain. That is why the translation in the phase-space
ensemble approach is not given by the momentum functional but by (\ref{translation}),
and the same is true for the other generators of the Galilei algebra.
This is especially important for the Hamiltonian functional (\ref{Hphasespace}).
Here we are accepting that the generator of time evolution is not
the energy observable, and we will have some more words about the
topic later on.

\subsection{From phase space back to configuration space}

We now face the problem of relating the ensembles in phase space with
the ensembles in configuration space. The procedure we will show is
based on the work of Klein \cite{hamilton-jacobi2}. First, Let us
recall that configuration space is a so-called Lagrangian submanifold
of the phase space, and the rules to work with these submanifolds
have been investigated for a long time \cite{mukunda,hamilton-jacobi}.
The entire idea here is to project the probability density and the
phase-space action into the configuration space in such a way that
their dependence on the momentum variable is eliminated.

We will proceed as follows: first notice that, by their definition
as probability densities, we have that

\begin{equation}
\rho(\mathbf{q})=\int d\mathbf{p}\,\varrho(\mathbf{q},\mathbf{p}),\label{roro}
\end{equation}
where $\rho(\mathbf{q})$ is positive and normalized in configuration
space. Second, let us recall that in configuration space the momentum
is related to a generating function $S(\mathbf{q},t)$ via

\begin{equation}
\mathbf{p}=\nabla S(\mathbf{q},\alpha,t),\label{generatingS}
\end{equation}
where $\alpha$ is a set of numbers that parametrize the Lagrangian
submanifold, the configuration space in this case. Now, the projection
to configuration space is realized by the following replacements

\begin{align}
\varrho(\mathbf{q},\mathbf{p},t) & \rightarrow\rho(\mathbf{q},t)\delta(\mathbf{p}-\nabla S),\label{Rro}\\
\mathcal{S}(\mathbf{q},\mathbf{p},t) & \rightarrow\mathcal{S}(\mathbf{q},\nabla S,t).\label{RS}
\end{align}

It is not immediately obvious, but both the projected action in phase
space $\mathcal{S}(\mathbf{q},\nabla S,t)$ and the generating function
of Eq (\ref{generatingS}) obey the Hamilton-Jacobi equation \cite{hamilton-jacobi2},
so they differ at most by a constant of motion

\[
\mathcal{S}(\mathbf{q},\nabla S,t)=S(\mathbf{q},t)+constant.
\]
We will ignore this constant from now on, so we can directly make
the replacement

\begin{equation}
\mathcal{S}(\mathbf{q},\mathbf{p},t)\rightarrow S(\mathbf{q},t).\label{SS}
\end{equation}

As mentioned, replacing (\ref{SS}) and (\ref{generatingS}) into
(\ref{Sl}) leads to the Hamilton-Jacobi equation. On the other hand,
assuming non-singular behavior of the quantities involves, replacing
(\ref{Rro}) into the Liouville equation (\ref{ro2}) and integrating
the momentum coordinates give the continuity equation in the configuration
space (\ref{conservation}).

The replacements (\ref{Rro}) and (\ref{RS}) also leads from the
Hamiltonian functional in phase-space (\ref{Hphasespace}) to the
Hamiltonian functional in configuration space (\ref{HC}), where we
only need to integrate out the momentum variables.

A similar situation occurs with the other generators of the Galilei
group, the phase-space generators become the ones given in configuration
space. In particular, we can see that the momentum observable merges
with the position translation functional (and the same is true for
the angular momentum and the dynamic mass moment).

\section{Quantum-classical hybrids}

We will construct in this section a hybrid theory of a classical
particle described in phase space interacting with a quantum particle
described in its configuration space. The quantum particle will have
mass $m_{1}$ and coordinate $\mathbf{x}$, and we will denote the
mass of the classical particle by $m_{2}$ and its phase-space coordinates
by $(\mathbf{q},\mathbf{p})$. For this composite system with probability
density $\varrho(\mathbf{q},\mathbf{p},\mathbf{x})$ and phase variable
$\mathcal{S}(\mathbf{q},\mathbf{p},\mathbf{x})$, we can write the
following basic observables

\begin{align}
P_{j} & =\int d\omega d\mathbf{x}\,\mathcal{\varrho}\left(p_{j}+\frac{\partial\mathcal{S}}{\partial x_{j}}\right),\nonumber \\
Q_{i} & =\int d\omega d\mathbf{x}\,\mathcal{\varrho}q_{i},\;\;X_{i}=\int d\omega d\mathbf{x}\,\mathcal{\varrho}x_{i}.
\end{align}

The generators of space-time transformations are, again, the composition
of the individual generator for each particle

\begin{align}
\Pi_{j}= & \int d\omega d\mathbf{x}\,\mathcal{\varrho}\left(\frac{\partial\mathcal{S}}{\partial q_{j}}+\frac{\partial\mathcal{S}}{\partial x_{j}}\right),\nonumber \\
L_{i}= & \int d\omega d\mathbf{x}\,\mathcal{\varrho}\varepsilon_{ijk}\left(q_{j}\frac{\partial\mathcal{S}}{\partial q_{k}}+p_{j}\frac{\partial\mathcal{S}}{\partial p_{k}}+x_{j}\frac{\partial\mathcal{S}}{\partial x_{k}}\right),\nonumber \\
G_{i}= & \int d\omega d\mathbf{x}\,\mathcal{\varrho}\left(m(q_{i}+x_{i})-m_{2}\frac{\partial\mathcal{S}}{\partial p_{i}}\right.\nonumber \\
 & \left.-t\left(\frac{\partial\mathcal{S}}{\partial q_{i}}+\frac{\partial\mathcal{S}}{\partial x_{i}}\right)\right).
\end{align}
The free-particle Hamiltonian functional is

\begin{align}
\mathcal{H}_{0}= & \int d\omega d\mathbf{x}\,\left[\mathcal{\varrho}\frac{\left|\nabla_{x}\mathcal{S}\right|^{2}}{2m_{1}}+\frac{\hbar^{2}}{2m_{1}}\frac{\left|\nabla\mathcal{\varrho}\right|^{2}}{\mathcal{\varrho}}\right.\nonumber \\
 & \left.-\mathcal{\varrho}\frac{\mathbf{p}^{2}}{2m_{2}}+\frac{1}{2}\left(\mathcal{\varrho}\nabla_{q}\mathcal{S}-\mathcal{S}\nabla_{q}\mathcal{\varrho}\right)\cdot\frac{\mathbf{p}}{m_{2}}\right]
\end{align}

Now, for this kind of hybrid system, $\mathcal{H}_{int}$ is not
as simple as in the previous examples. The reason is that the Galilean
covariance of the theory does not guarantee the conservation of the
physical quantities of interest. For example, the following functional
is Galilei invariant
\[
\Phi=\int d\omega_{1}d\mathbf{x}\:\rho V(\left|\mathbf{q}-\mathbf{x}\right|).
\]
However, $\Phi$ does not conserve the total momentum
\[
\left\{ \Phi,P_{j}\right\} _{f}\neq0.
\]
It is also unacceptable to have interaction terms that conserve the
total momentum but that are not invariant under Galilei transformations.

Fortunately, it is possible to have interaction functionals that are
both translational invariant and that conserve momentum. We are not
going to try to classify all the possibilities, but we mention the
following example

\begin{align}
\mathcal{H}_{int}= & \Phi+\frac{1}{2}\int d\omega d\mathbf{x}\,\nonumber \\
 & \times\left(\mathcal{S}\nabla_{p}\mathcal{\varrho}-\mathcal{\varrho}\nabla_{p}\mathcal{S}\right)\cdot\nabla_{q}V(\left|\mathbf{q}-\mathbf{x}\right|),\nonumber \\
\left\{ \mathcal{H}_{int},P_{j}\right\}  & =\left\{ \mathcal{H}_{int},\Pi_{j}\right\} =0.
\end{align}

Using the full Hamiltonian

\[
\mathcal{H}=\mathcal{H}_{0}+\mathcal{H}_{int},
\]
we arrive at the following (Galilei covariant) equations of motion

\begin{align}
\frac{\partial\mathcal{\varrho}}{\partial t}+\frac{\nabla_{x}\cdot\left(\mathcal{\varrho}\nabla_{x}\mathcal{S}\right)}{m_{1}} & =\nabla_{p}\mathcal{\varrho}\cdot\nabla_{q}V-\frac{\mathbf{p}\cdot\nabla_{q}\mathcal{\varrho}}{m_{2}},\nonumber \\
\frac{\partial\mathcal{S}}{\partial t}+\frac{\left|\nabla_{x}\mathcal{S}\right|^{2}}{2m_{1}}= & -\frac{\hbar^{2}}{2m_{1}}\frac{\nabla^{2}\sqrt{\mathcal{\varrho}}}{\sqrt{\mathcal{\varrho}}}+\frac{\mathbf{p}^{2}}{2m_{2}}-V\nonumber \\
 & +\nabla_{p}\mathcal{S}\cdot\nabla_{q}V-\frac{\mathbf{p}\cdot\nabla_{q}\mathcal{S}}{m_{2}}.\label{tronci}
\end{align}

The equations (\ref{tronci}) were first written in \cite{tronci1}.
They are the Madelung form of the wave-function hybrid theory originating
from a partial geometric quantization of the Koopman-van Hove equation
\cite{DGBT,tronci2,tronci3}.

Finally, we can ask if Eqs (\ref{tronci}) can be projected into the
configuration space of the classical particle. I cannot give a satisfactory
answer to this question at this point. However, notice that if a procedure
similar to the one given in section 3.2 is correct, namely, if we
can make the substitutions

\begin{align}
\mathbf{p} & \rightarrow\nabla_{q}S(\mathbf{q},\mathbf{x},t),\nonumber \\
\varrho(\mathbf{q},\mathbf{p},\mathbf{x},t) & \rightarrow\rho(\mathbf{q},\mathbf{x},t)\delta(\mathbf{p}-\nabla_{q}S),\nonumber \\
\mathcal{S}(\mathbf{q},\mathbf{p},\mathbf{x},t) & \rightarrow\mathcal{S}(\mathbf{q},\mathbf{x},\nabla_{q}S,t)\rightarrow S(\mathbf{q},\mathbf{x},t),\label{RS-1}
\end{align}
then Eqs (\ref{tronci}) reduce to the hybrid equations of Hall and
Reginatto \cite{reginatto2}

\begin{align}
\frac{\partial S}{\partial t}+\frac{\left|\nabla_{x}S\right|^{2}}{2m_{1}}+\frac{\left|\nabla_{q}S\right|^{2}}{2m_{2}}+V & =\frac{\hbar^{2}}{2m_{1}}\frac{\nabla^{2}\sqrt{\rho}}{\sqrt{\rho}},\nonumber \\
\frac{\partial\rho}{\partial t}+\frac{\nabla_{x}\cdot\left(\mathcal{\rho}\nabla_{x}S\right)}{m_{1}}+\frac{\nabla_{q}\cdot\left(\rho\nabla_{q}S\right)}{m_{2}} & =0.
\end{align}

\section{Discussion and final remarks}

The close relationship between the hybrid model originating in ensembles
in configuration space with the one that comes from the Koopman-von
Hove equation requires further study. In particular, our results seem
to be at odds with the conclusions presented in \cite{entanglement}.
The question investigated in \cite{entanglement} is ``Can quantum
systems become entangled via a classical intermediary?''. The conclusion
offered is that hybrids based on ensembles in configuration space
can produce entanglement, while Koopman hybrids can not. However,
this last statement is based on the incorrect premise that only commuting
operators are used for the classical particle. Indeed, as we can see
in (\ref{tronci}), in the hybrid theory the (commuting) position
and momentum of the classical particle appear together with the derivative
operators $\nabla_{p}$ and $\nabla_{q}$. 

It has to be mentioned that the equations presented here are not entirely
equivalent to the Hybrid system developed in \cite{DGBT} because
the relationship between the KvH wavefunction and the probability
density is different. Instead of using $\psi(\mathbf{q},\mathbf{p})=\sqrt{\varrho}e^{i\mathcal{S}/\hbar}$
that leads to $\varrho=\left|\psi\right|^{2}$, the authors of \cite{DGBT}
use a more involved formula. The reason of the discrepancy is that
in \cite{DGBT} the authors make the time evolution functional and
the energy functional to coincide, and, as we mentioned sections 3
and 3.1, that is not the case in the formalism developed here. Further
work is required to see if we can make the two approaches to be fully
compatible. 

Lastly, whether the replacement of Eq (\ref{RS-1}) is correct remains
to be seen. If so, this will imply some sort of equivalence between
the two hybrid models. However, it can be the case that we are not
allowed to do the projection to configuration space, or that it only
can be done under certain circumstances or only for some special cases.
This issue will be investigated in future work.


\begin{thebibliography}{10}
\bibitem{barcelo}C. Barcel\'o, R. Carballo-Rubio, L. J. Garay, and
R. G\'omez- Escalante, Phys. Rev. A $\mathbf{86}$, 042120 (2012).

\bibitem{sudarshan1}E. C. G. Sudarshan, Pramana $\mathbf{6}$, 117
(1976).

\bibitem{sudarshan2}T. N. Sherry and E. C. G. Sudarshan, Phys. Rev.
D $\mathbf{18}$, 4580 (1978); $\mathbf{20}$, 857 (1979); S. R. Gautam,
T. N. Sherry, and E. C. G. Sudarshan, ibid.$\mathbf{20}$, 3081 (1979).

\bibitem{gravity}Boucher W and Traschen J 1988 Semiclassical physics
and quantum fluctuations Phys. Rev. D $\mathbf{37}$ 3522--32.

\bibitem{koopman}B. O. Koopman, Proc. Natl. Acad. Sci. USA $\mathbf{17}$,
315 (1931); J. von Neumann, Ann. Math. $\mathbf{33}$, 587 (1932);
,$\mathbf{33}$, 789 (1932).

\bibitem{koopman2}M. Radonji\'{c}, D. B. Popovi\'{c}, S. Prvanovi\'{c},
and N. Buri\'{c} Phys. Rev. A $\mathbf{89}$, 024104 (2014).

\bibitem{koopman3}D. Chru\'{s}ci\'{n}ski, A. Kossakowski, G. Marmo
and E. C. G. Sudarshan, Open Syst. Inf. Dyn. $\mathbf{18}$ 339--51,
(2011) 

\bibitem{terno1}A. Peres and D. R. Terno, Phys. Rev. A $\mathbf{63}$,
022101 (2001). 

\bibitem{terno2}D. R. Terno, Found. Phys. $\mathbf{36}$, 102 (2006). 

\bibitem{salcedo1}L. L. Salcedo, Phys. Rev. A $\mathbf{54}$, 3657
(1996). 

\bibitem{salcedo2}L. L. Salcedo, Phys. Rev. A $\mathbf{85}$, 022127
(2012).

\bibitem{galilean}A. D. Berm\'udez Manjarres and N. Mar\'in-Medina, Phys.
Rev. A. $\mathbf{102}$, 042221 (2020).

\bibitem{tronci1}F. Gay-Balmaz and C. Tronci, Nonlinearity $\mathbf{33}$
5383 (2020).

\bibitem{DGBT}D. I. Bondar , F. Gay-Balmaz and C. Tronci, Proc. R.
Soc. A $\mathbf{475}$ 20180879 (2019)

\bibitem{tronci2}F. Gay-Balmaz and C. Tronci, ``From quantum hydrodynamics
to Koopman wavefunctions I Geometric Science of Information (Lecture
Notes in Computer Science vol 12829) ed F Nielsen and F Barbaresco''
(Berlin: Springer, 2021) 

\bibitem{tronci3}F. Gay-Balmaz and C. Tronci, ``From quantum hydrodynamics
to Koopman wavefunctions II Geometric Science of Information (Lecture
Notes in Computer Science vol 12829) ed F Nielsen and F Barbaresco''
(Berlin: Springer, 2021)

\bibitem{reginato}M. J. W. Hall and M. Reginatto, ``Ensembles on
Configuration Space: Classical, Quantum, and Beyond'' (Switzerland:
Springer International Publishing, 2016).

\bibitem{reginatto2}M. J. W. Hall and M. Reginatto, Phys. Rev. A
$\mathbf{72}$, 062109 (2005).

\bibitem{reginatto3}M. Reginatto, J. Phys. Conf. Ser. $\mathbf{1612}$
012023 (2020).

\bibitem{klein} U. Klein, Quantum Stud.: Math. Found. $\mathbf{5}$
219--27 (2018)

\bibitem{me1}A. D. Berm\'udez Manjarres, M. Nowakowski and D. Batic,
Ann. Phys (NY), $\mathbf{416}$ 168157 (2020).

\bibitem{m2}A. D. Berm\'udez Manjarres, Ann. Phys (NY) $\mathbf{431}$
168539 (2021).

\bibitem{GKVN1}A. Loinger Ann. Phys (NY) $\mathbf{20}$ 132 (1962).

\bibitem{GKVN2}A. Loinger Ann. Phys (NY) $\mathbf{23}$ 23 (1963)
.

\bibitem{GKVN3}P. Gulmanelli Phys. Lett. $\mathbf{5}$320 (1963).

\bibitem{GKVN4}G. Lugarini and M. Pauri Ann. Phys., NY $\mathbf{38}$
299--314 (1966).

\bibitem{m3}A. D. Bermudez Manjarres, J. Phys. A: Math. Theor. $\mathbf{54}$
444001 (2021).

\bibitem{mukunda}N. Mukunda, Proc. Indian Acad. Sci. $\mathbf{87}$
85--105 6 (1978)

\bibitem{hamilton-jacobi}A. Carosso, ``Geometric quantization''
(2018) arXiv:1801.02307{[}math-ph{]}

\bibitem{hamilton-jacobi2}U. Klein, arXiv:2202.13356v1

\bibitem{entanglement}M. J. W. Hall and M. Reginatto, J. Phys. A:
Math. Theor. $\mathbf{51}$ (2018) 085303 

\bibitem{bondar}Bondar D I, Cabrera R, Lompay R R, Ivanov M Y and
Rabitz H A 2012 Phys. Rev. Lett. 109 190403
\end{thebibliography}
\end{document}